\documentclass{ws-procs9x6}

\def\etal {{\it et al.}}

\begin{document}

\title{SEARCH FOR LORENTZ VIOLATION IN A\\
SHORT-RANGE GRAVITY EXPERIMENT}

\author{D.\ BENNETT, V.\ SKAVYSH, AND J.\ LONG$^*$}

\address{Physics Department, Indiana University \\
727 E.\ Third St., Bloomington, IN 47408, USA\\
$^*$E-mail: jcl@indiana.edu}

\begin{abstract}
An experimental test of the Newtonian inverse square law at short
range has been used to set limits on Lorentz violation in
the pure gravity sector of the Standard-Model Extension. On account of
the planar test mass geometry, nominally null with respect to   
$1/r^{2}$ forces, the limits derived for the SME coefficients of
Lorentz violation are on the order $\bar{s}^{JK}\sim 10^{4}$.   
\end{abstract}

\bodymatter

\section{Introduction}
Local Lorentz invariance is at the foundation of both
the Standard Model of particle physics and General Relativity, but is
generally poorly tested for the latter theory. Violation of Lorentz 
symmetry would break the isotropy of spacetime, permitting the vacuum
to fill with `background' fields that have a preferred direction.
Interaction of the test masses in a terrestrial gravity
experiment with these fields could result in a sidereal modulation of
the force between the masses, thus providing a test of Lorentz
invariance in gravity.  

A quantitative description of Lorentz violation consistent with local
field theory is given by the Standard-Model Extension (SME), which has
recently been expanded to include gravitational effects.\cite{Kostelecky04}  The
expression for the potential between two test masses $m_{1}$ and
$m_{2}$ in the `pure gravity' sector of the SME is given by: 
\begin{equation}
V = -\frac{Gm_{1}m_{2}}{|\vec{x}_{1}-\vec{x}_{2}|}\left(1+\frac{1}{2}\hat{x}^{\hat{j}}\hat{x}^{\hat{k}}\bar{s}^{\hat{j}\hat{k}}\right), 
\label{eq:LVgrav}
\end{equation}
where $\vec{x}_{1}-\vec{x}_{2}$ is the vector separating $m_{1}$ and
$m_{2}$, $\hat{x}^{\hat{i}}$ is the projection of the unit vector
along $\vec{x}_{1}-\vec{x}_{2}$ in the $i$th direction, and
$\bar{s}^{\hat{j}\hat{k}}$ is a set of 9 dimensionless coefficients
for Lorentz violation in the standard laboratory frame.\cite{Bailey06}
Motivated by Ref.\ \refcite{Bailey06}, this report presents a test of
Eq.~\eqref{eq:LVgrav} in a laboratory gravity experiment.

\section{The Indiana short--range experiment}
The Indiana experiment is optimized for sensitivity to possible new forces in nature at short range, which in turn could
arise from new elementary particles or even extra spacetime
dimensions.\cite{Adelberger03,Dimopoulos03} It is described in
detail elsewhere;\cite{Long03_2,Long03_3,Jensen07} here we concentrate
on the essential features.    
To attain sensitivity at short range, the test
masses must in general be scaled to that range, so as not to be
overwhelmed by Newtonian forces at larger scales.  

\begin{figure}[htbp]
\begin{center}
\psfig{file=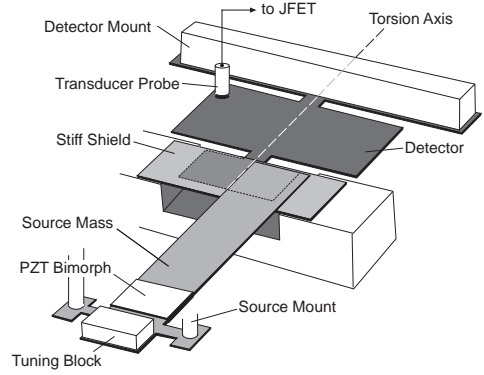,width=2.5in}
\end{center}
\caption{Central components of Indiana short--range
  experiment, to scale (the long dimension of the detector mount is
  approximately 2.5~cm.)} 
\label{fig:expt}  
\end{figure}

The experiment is illustrated in Fig.~\ref{fig:expt}. The test masses
consist of 250~$\mu$m thick planar tungsten 
oscillators, separated by a gap of 100~$\mu$m, with a stiff conducting
shield in between them to suppress 
electrostatic and acoustic backgrounds.  The planar geometry
concentrates as much mass as possible at
the scale of interest, and is nominally null with
respect to $1/r^{2}$ forces.  This is effective in suppressing the
Newtonian background relative to new short-range effects, but has
detrimental consequences for testing the pure gravity sector of the
SME.  The force--sensitive `detector' mass is driven by the
force--generating `source' mass at a resonance near 1~kHz, placing a
heavy burden on
vibration isolation.  The 1 kHz operation is chosen since at this
frequency it is possible to construct a simple, passive vibration
isolation system.  The entire apparatus is enclosed in a vacuum
chamber and operated at $10^{-7}$ torr to further reduce the acoustic
coupling.  Detector oscillations are read out with a capacitive
transducer and JFET amplifier.

This design has proven effective for suppressing all background forces
to the extent that the only
effect observed is thermal noise due to dissipation in the detector
mass.  After a run over the course of several days in 2002, this
experiment set the strongest limits on new forces of nature between 10
and 100~$\mu$m.\cite{Long03_3}  The experiment has since been optimized to
explore gaps below 50~$\mu$m; operation at the thermal noise limit has
recently been demonstrated but with limited statistics.

\section{Lorentz violating force for planar geometry}
Analysis of the 2002 data for evidence of Lorentz violation
requires a theoretical expression for the Lorentz violating force for
the particular test mass geometry. The $j$th
component of the differential force corresponding to the modified
potential in Eq.~\eqref{eq:LVgrav} is given by:
\begin{equation}
dF^{\hat{j}}=Gdm_{1}dm_{2}\left(-\frac{x^{\hat{j}}}{x^{3}}-\frac{3}{2}\frac{x^{\hat{j}}}{x^{5}}x^{\hat{j}}x^{\hat{k}}\bar{s}^{\hat{i}\hat{k}}+\frac{x^{\hat{k}}\bar{s}^{\hat{j}\hat{k}}}{x^{3}}\right),
\label{eq:LVdF}
\end{equation}
where $x^{\hat{j}}$ and $x^{\hat{k}}$ can be understood as
components of the unit vector pointing from a differential mass element
$dm_{1}$ in the source mass to a corresponding
element $dm_{2}$ in the detector.  The SME coefficients in the lab
frame $\bar{s}^{\hat{j}\hat{k}}$ are related to the coefficients in
the Sun-centered celestial equatorial frame $\bar{s}^{JK}$ by
$\bar{s}^{\hat{j}\hat{k}} = R^{T}\bar{s}^{JK}R$, where the matrix $R$
is given by Eq.~(C1) of Ref.\ \refcite{Kostelecky02}.

The detector is a constrained mechanical oscillator with distributed
mass, so the experiment is sensitive to the force density averaged over
the free modal displacement:
\begin{equation}
F=\frac{1}{|z_{\mbox{max}}|}\int_{D}d^{3}\vec{x}\vec{z}^{F}(\vec{x})\cdot\vec{f}(\vec{x}).  
\label{eq:LVFavg}
\end{equation}
Here, the modal displacement $\vec{z}^{F}(\vec{x})$ is
derived from a finite-element model of the detector mass with overall
normalization factor
$|z_{\mbox{max}}|$, and the components of the force density
$\vec{f}(\vec{x})$ are given by Eq.~\eqref{eq:LVdF} above.  

Equation \eqref{eq:LVFavg} is evaluated by Monte Carlo integration with the
SME coefficients expressed in the Sun-centered frame and the
geometrical parameters listed in Table 2 of Ref.\ \refcite{Long03_2}.
As the experiment is done on resonance, the Monte Carlo algorithm
computes the Fourier amplitude of Eq.~\eqref{eq:LVFavg} averaged over a
complete cycle of the source mass oscillation, taking into account the
measured source mass curvature and
mode shape.  The result can be expressed as a Fourier series of the
time dependence,
\begin{equation}
F=C_{1}+C_{2}\sin(\omega_{\oplus}T)+C_{3}\cos(\omega_{\oplus}T)+C_{4}\sin(2\omega_{\oplus}T)+C_{5}\cos(2\omega_{\oplus}T),  
\label{eq:LVF}
\end{equation}
where $\omega_{\oplus}$ is the Earth's sidereal rotation frequency,
and the time $T$ is measured in the Sun-centered frame. The $C_{k}$
are functions of the SME coefficients and test mass geometry (and the
laboratory colatitude angle) and are given explicitly in
Table~\ref{tab:sig}.
The statistical errors from the Monte Carlo on the numerical
coefficient of each $\bar{s}^{JK}$ term in Table~\ref{tab:sig} are at
most a few percent and ignored for the purpose of computing
preliminary limits on the SME coefficients.

\begin{table}
\tbl{Signal components for the Indiana short range gravity experiment
  (when located at colatitude $\chi=0.87$ in 2002).} 
{\begin{tabular}{@{}cr@{}}
\toprule
\toprule
\multicolumn{1}{c}{Component} & \multicolumn{1}{c}{Amplitude ($\times 10^{-16}$~ N)}\\
\colrule
$C_{1}$ &
  $1.12\bar{s}^{XX}+0.00\bar{s}^{XY}-7.78\bar{s}^{XZ}-3.48\bar{s}^{YY}+0.00\bar{s}^{YZ}-0.21\bar{s}^{ZZ}$\\
$C_{2}$ &
  $0.07\bar{s}^{XX}+0.76\bar{s}^{XY}+0.02\bar{s}^{XZ}+0.00\bar{s}^{YY}-0.64\bar{s}^{YZ}-0.07\bar{s}^{ZZ}$\\
$C_{3}$ &
  $-0.49\bar{s}^{XX}+0.11\bar{s}^{XY}-0.17\bar{s}^{XZ}+0.00\bar{s}^{YY}-0.09\bar{s}^{YZ}+0.49\bar{s}^{ZZ}$\\
$C_{4}$ &
  $0.06\bar{s}^{XX}-0.08\bar{s}^{XY}+0.15\bar{s}^{XZ}-0.16\bar{s}^{YY}-0.10\bar{s}^{YZ}-0.09\bar{s}^{ZZ}$\\
$C_{5}$ &
  $0.03\bar{s}^{XX}+0.20\bar{s}^{XY}+0.06\bar{s}^{XZ}-0.06\bar{s}^{YY}+0.24\bar{s}^{YZ}+0.04\bar{s}^{ZZ}$\\
\botrule
\end{tabular}}
\label{tab:sig}
\end{table}
 
\section{Limits on coefficients for Lorentz violation }
A preliminary analysis of the 2002 data set (incorporating statistical
errors only) for signals of Lorentz
violation has been completed, following Ref.\ \refcite{Chung09}. In this
analysis, the time stamps in the data are extracted and offset
relative to the effective $T_{0}$ in the Sun-centered frame (solar
noon on the 2002 vernal equinox).  The results are shown in
Fig.~2 of Ref.\ \refcite{Jensen07}. The discrete Fourier transforms of
the data at each frequency component of the signal ($0, \omega_{\oplus},
2\omega_{\oplus}$) are computed, with errors.  Gaussian probability
distributions for the experiment at 
each signal frequency component are constructed, using the difference
between the Fourier transforms and the predicted signals
(Table~\ref{tab:sig}) as the means.  As in Ref.\ \refcite{Chung09},
the overlap of the Fourier components characteristic of the finite data
set (Fig.~2 of Ref.\ \refcite{Jensen07}) is quantified by scaling each
term of the predicted 
signal with the element of a covariance matrix formed from the time
stamps of the data.  Finally, an overall probability distribution for
the experiment is constructed from the product of the component
distributions.  Means and errors of particular
$\bar{s}^{JK}$ (for example, $\bar{s}^{XX}$) are then computed by
integration of the distribution over all $\bar{s}^{JK}$
except $\bar{s}^{XX}$.  Results for each $\bar{s}^{JK}$ (except one)
are shown in Table~\ref{tab:results}.  
\begin{table}
\tbl{Limits on pure-gravity sector SME Lorentz violation coefficients
  from the Indiana short-range gravity experiment.} 
{\begin{tabular}{@{}cr@{}}
\toprule
\toprule
\multicolumn{1}{c}{Coefficient} & \multicolumn{1}{r}{Mean and error (2$\sigma$)}\\
\colrule
$\bar{s}^{XX}$ & \hphantom{0000000000000000000000000000}$(-0.04\pm 4.90)\times 10^{4}$\\
$\bar{s}^{XY}$ & $(-0.07\pm 6.12)\times 10^{4}$\\
$\bar{s}^{XZ}$ & $(-0.01\pm 2.56)\times 10^{3}$\\
$\bar{s}^{YZ}$ & $(-0.06\pm 5.83)\times 10^{4}$\\
$\bar{s}^{ZZ}$ & $(0.08\pm 6.68)\times 10^{4}$\\
\botrule
\end{tabular}}
\label{tab:results}
\end{table}
This procedure is sufficient to determine the SME coefficients and
account for their correlations, as long as the number of signal
components equals or exceeds the number of coefficients in
Table~\ref{tab:sig}, which is not the case; for this analysis,
$\bar{s}^{YY}$, to which the experiment is apparently least
sensitive, has been set to zero.

The limits are on the order of $10^{3}$-$10^{4}$. This is as expected,
since, as pointed out in Ref.\ \refcite{Bailey06}, Lorentz violation
leads to a misalignment of the force associated with
Eq.~\eqref{eq:LVgrav} relative to the vector $\vec{x}_{1}-\vec{x}_{2}$,
however the
$1/r^{2}$ behavior of the force is preserved.  The planar test
mass geometry renders the Indiana experiment insensitive to $1/r^{2}$
forces (about $10^{3}$ times gravitational strength at
10~$\mu$m) which is directly reflected in the limits on the SME
coefficients.

\section*{Acknowledgments}

The authors would like to thank H. Muller for useful discussions of
Ref.\ \refcite{Chung09}.


\begin{thebibliography}{xx}

\bibitem{Kostelecky04}
V.A.\ Kosteleck\'y, 
Phys.\ Rev.\ D {\bf 69}, 105009 (2004).

\bibitem{Bailey06}
Q.G.\ Bailey and V.A.\ Kosteleck\'y, 
Phys.\ Rev.\ D {\bf 74}, 045001 (2006).

\bibitem{Adelberger03}
E.\ Adelberger \etal, 
Ann.\ Rev.\ Nucl.\ Part.\ Sci.\ {\bf 53}, 77 (2003).

\bibitem{Dimopoulos03} 
S.\ Dimopoulos and A.A.\ Gearci, 
Phys.\ Rev.\ D {\bf 68}, 124021 (2003).

\bibitem{Long03_2}
J.C.\ Long \etal, 
arXiv:hep-ph/0210004.

\bibitem{Long03_3}
J.C.\ Long \etal, 
Nature {\bf 421}, 922 (2003).

\bibitem{Jensen07}
W.A.\ Jensen, S.M.\ Lewis, and J.C.\ Long,
in V. A.\ Kosteleck\'y, ed.,  
{\it Proceedings of the Fourth Meeting on CPT and Lorentz Symmetry,}
World Scientific, Singapore, 2008.

\bibitem{Kostelecky02} 
V.A.\ Kosteleck\'y and M.\ Mewes, 
Phys.\ Rev.\ D {\bf 66}, 056005 (2002).

\bibitem{Chung09}
K.-Y.\ Chung \etal, 
Phys.\ Rev.\ D {\bf 80}, 016002 (2009).

\end{thebibliography}
\end{document}